\documentclass[
 reprint,
 superscriptaddress,
 amsmath,amssymb,
 aps,
 prl,
]{revtex4-2}

\usepackage{graphicx}
\usepackage{dcolumn}
\usepackage{bm}

\usepackage{soul}
\usepackage{xcolor}
\usepackage{graphicx}

\begin{document}

\title{\boldmath Axion Dark Matter Search with Near-KSVZ Sensitivity Using the TM$_{020}$ Mode}

\author{Sungjae Bae}
\thanks{These authors contributed equally to this work.}
\affiliation{Dark Matter Axion Group, Institute for Basic Science, Daejeon 34126 Republic of Korea}
\affiliation{Center for Axion and Precision Physics Research, Institute for Basic Science, Daejeon 34051, Republic of Korea}
\affiliation{Department of Physics, Korea Advanced Institute of Science and Technology, Daejeon 34141, Republic of Korea}
\author{Junu Jeong}
\thanks{These authors contributed equally to this work.}
\affiliation{Center for Axion and Precision Physics Research, Institute for Basic Science, Daejeon 34051, Republic of Korea}
\affiliation{Oskar Klein Centre, Department of Physics, Stockholm University, AlbaNova, SE-10691 Stockholm, Sweden}
\author{Younggeun Kim}
\thanks{These authors contributed equally to this work.}
\affiliation{Center for Axion and Precision Physics Research, Institute for Basic Science, Daejeon 34051, Republic of Korea}
\affiliation{Johannes Gutenberg-Universit{\"a}t Mainz, 55122 Mainz, Germany}
\affiliation{GSI Helmholtzzentrum für Schwerionenforschung GmbH, 64291 Darmstadt, Germany}
\author{SungWoo Youn}
\thanks{swyoun@ibs.re.kr}
\affiliation{Dark Matter Axion Group, Institute for Basic Science, Daejeon 34126 Republic of Korea}
\affiliation{Center for Axion and Precision Physics Research, Institute for Basic Science, Daejeon 34051, Republic of Korea}
\author{Jinsu Kim}
\affiliation{Dark Matter Axion Group, Institute for Basic Science, Daejeon 34126 Republic of Korea}
\affiliation{Center for Axion and Precision Physics Research, Institute for Basic Science, Daejeon 34051, Republic of Korea}
\author{Arjan F. van Loo}
\affiliation{RIKEN Center for Quantum Computing, Wako, Saitama 351-0198, Japan}
\affiliation{Department of Applied Physics, Graduate School of Engineering, The University of Tokyo, Bunkyo-ku, Tokyo 113-8656, Japan}
\author{Yasunobu Nakamura}
\affiliation{RIKEN Center for Quantum Computing, Wako, Saitama 351-0198, Japan}
\affiliation{Department of Applied Physics, Graduate School of Engineering, The University of Tokyo, Bunkyo-ku, Tokyo 113-8656, Japan}
\author{Seonjeong Oh}
\affiliation{Center for Axion and Precision Physics Research, Institute for Basic Science, Daejeon 34051, Republic of Korea}
\author{Taehyeon Seong}
\affiliation{Center for Axion and Precision Physics Research, Institute for Basic Science, Daejeon 34051, Republic of Korea}
\author{Sergey Uchaikin}
\affiliation{Center for Axion and Precision Physics Research, Institute for Basic Science, Daejeon 34051, Republic of Korea}
\author{Jihn E. Kim}
\affiliation{Department of Physics, Seoul National University, Seoul 08826, Republic of Korea}
\author{Yannis K. Semertzidis}
\affiliation{Center for Axion and Precision Physics Research, Institute for Basic Science, Daejeon 34051, Republic of Korea}
\affiliation{Department of Physics, Korea Advanced Institute of Science and Technology, Daejeon 34141, Republic of Korea}

\date{\today}

\begin{abstract}
Dark matter remains one of the most profound mysteries in modern physics, with axions, a hypothetical particle proposed to resolve the strong CP problem, standing as a compelling candidate. 
Among various experimental strategies, cavity haloscopes currently offer the most sensitive method to detect axions, though their searches have largely been confined to axion masses below 10\,$\mu$eV. 
However, recent theoretical developments suggest that the axion mass lies beyond this range. 
Higher-order cavity modes have been explored as a methodological approach to expand the search range, albeit with limited success in achieving both high sensitivity and broad tunability.
In this work, we present a sensitive search for axions with masses around 21\,$\mu$eV, utilizing the TM$_{020}$ mode of a cylindrical cavity, which incorporated an innovative tuning mechanism.
Our results reached 1.7 times the KSVZ sensitivity over 100\,MHz, representing a significant improvement in this mass range and contributing to the experimental search for axion dark matter at higher masses.
\end{abstract}

\maketitle


The nature of dark matter remains one of the greatest mysteries in modern science, and its discovery could provide deep insights into fundamental physics. 
Among the various candidates~\cite{RevModPhys.90.045002}, the axion is particularly compelling, initially proposed to address the Charge-Parity (CP) problem in quantum chromodynamics (QCD). 
Predicted by the Peccei-Quinn mechanism through spontaneous breaking of global U(1) axial symmetry, the axion possesses well-defined properties~\cite{PecceiQuinn:PRL:1977}.
The invisible models, Kim-Shifman-Vainshtein-Zakharov (KSVZ)~\cite{Kim:PRL:1979,Shiftman:NPB:1980}
and Dine-Fischler-Srednicki-Zhitnitsky (DFSZ)~\cite{Zhitnitsky:SJNP:1980,Dine:PLB:1981}
, predict light, weakly interacting axions, suggesting they may have formed as cold, non-relativistic particles in the early Universe and could significantly contribute to the observed dark matter density~\cite{Preskill:PLB:1983,Abbott:PLB:1983,Dine:PLB:1983}.

Over the past few decades, experimental efforts to detect axion signals have significantly progressed
~\cite{cajohare:github:2020,Yannis:SciAdv:2022}.
Haloscope experiments, which exploit the electromagnetic interactions of axions in our galactic halo, employ high-quality resonant cavities immersed in strong magnetic fields to enhance the conversion of axions into detectable microwave photons~\cite{Sikivie:PRL:1983}.
In cavity haloscopes, the axion-to-photon conversion power is given by
\begin{widetext}
\begin{equation}
    P_{a\gamma} \simeq 9.2 \times 10^{-23}\,{\rm W} \left(\frac{g_{\gamma}}{-0.97}\right)^2 
    \left(\frac{\rho_a}{0.45\, {\rm GeV/cm^3}}\right) 
    \left(\frac{\nu_a}{5.1\,{\rm GHz}}\right) 
    \left(\frac{\langle \mathbf{B}_{e}^{2} \rangle}{\left(9.8\,{\rm T}\right)^2}\right)
    \left(\frac{V_c}{1.4\,{\rm L}}\right)
    \left(\frac{G}{0.5}\right) 
    \left(\frac{Q_c}{10^5}\right),
    \label{eq:conv_power}
\end{equation}
\end{widetext}
where $g_{\gamma}$ represents the axion-photon coupling coefficient with values of $-$0.97 and 0.36 for the KSVZ and DFSZ models, respectively, $\rho_{a}$ and $\nu_a(=m_a/2\pi)$ denote the local density and Compton frequency (mass) of the dark matter axion, $\langle \mathbf{B}_{e}^{2} \rangle$ is the average squared external magnetic field within the cavity volume $V_c$, and $Q_c(=Q_l (1+\beta))$ is the unloaded (loaded) cavity quality factor with $\beta$ being the antenna coupling coefficient.
The geometry factor $G$ refers to how efficiently a resonant cavity mode couples to the axion-induced electromagnetic signal:
\begin{equation}
\label{eq:geo_fac} 
    G = \frac{\left| \int \mathbf{E}_{r}\cdot \mathbf{B}_e \, dV_{c} \right|^{2}}{\int \epsilon \left| \mathbf{E}_{r} \right|^{2}dV_{c} \int \left| \mathbf{B}_e \right|^{2} dV_{c}},
\end{equation}
where $\mathbf{E}_{r}$ is the electric field of the mode and $\epsilon$ is the dielectric constant inside the cavity.

Although cavity haloscopes are among the most effective methods for detecting axion dark matter, current sensitive searches are primarily limited to low-mass regions, typically within several micro-electron volts, due to the practical constraints of existing experimental setups optimized for maximum detection volume. 
To address this, various advancements in resonator designs~\cite{jeong2020search} and new detection techniques~\cite{Caldwell2017PRL,lawson2019tunable}
have been explored to extend the sensitivity of axion searches into higher-mass regions.
In parallel, theoretical studies, particularly within post-inflationary scenarios, have constrained the plausible axion mass range based on various production mechanisms, such as misalignment and topological defects~\cite{Wantz:PRD:2010, Kawasaki:PRD:2015, Borsanyi:Nature:2016, Bonati:JHEP:2016, Klaer:JCAP:2017, Dine:PRD:2017, Buschmann:PRL:2020, Buschmann:NatureComm:2022,Kim:2024aa}, often complemented by astrophysical observations~\cite{PhysRevD.80.035024,PhysRevLett.113.011802}.
They generally suggest that axions may have a mass beyond the reach of current sensitive experiments.
Consequently, extending experimental searches to higher-mass regions is crucial for testing these theoretical frameworks and evaluating the viability of axions as dark matter.

Higher-order resonant modes present advantages for axion searches at high frequencies, naturally extending the search range without sacrificing cavity volume while providing higher quality factors, compared to the fundamental mode.
However, their practical application has been limited by small geometry factors and challenges in frequency tuning due to the complex field variations.
A recent study introduced an auxetic structure to enhance the tunability while preserving a suitable geometry factor for the TM$_{020}$ mode~\cite{bae2024search}.
Implemented in an axion search around 5.2\,GHz, the mechanism enabled a 100-MHz scan with improved geometry factors, demonstrating its effectiveness for higher-order mode tuning.
In this Letter, we report new results from a subsequent search achieving near-KSVZ sensitivity over an axion frequency range of 5.07--5.17\,GHz, further advancing the feasibility of higher-order mode-based axion searches.

\begin{figure}
\centering
\includegraphics[width=0.9\linewidth]{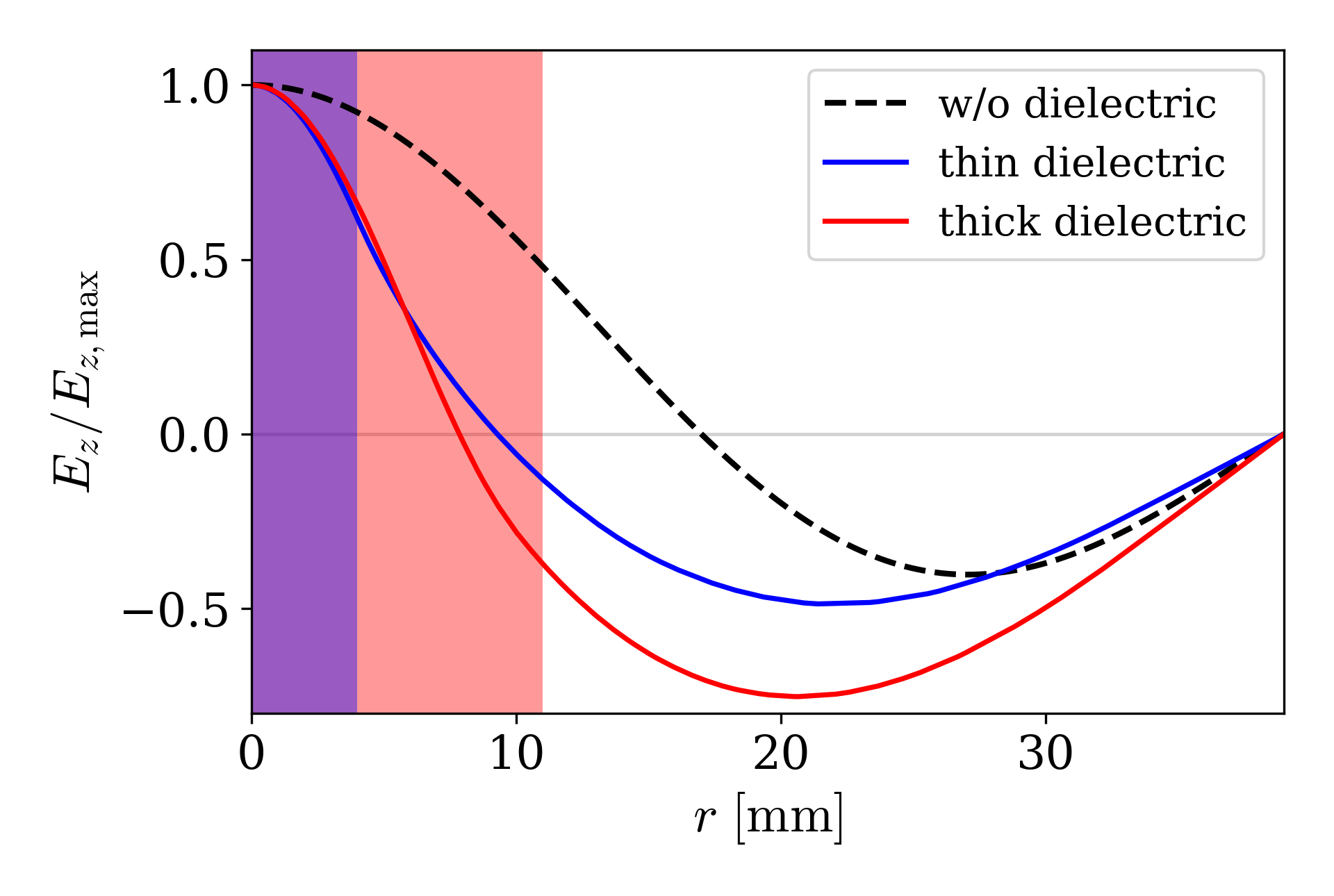}
\caption{Electric field profiles of the TM$_{020}$ mode without (dashed black line) and with (solid colored lines) a dielectric placed at the cavity center.
The thicknesses of the dielectric materials are indicated by the shaded regions.}
\label{fig:field_profile}
\end{figure}

\begin{figure}[b]
\centering
\includegraphics[width=1.0\linewidth]{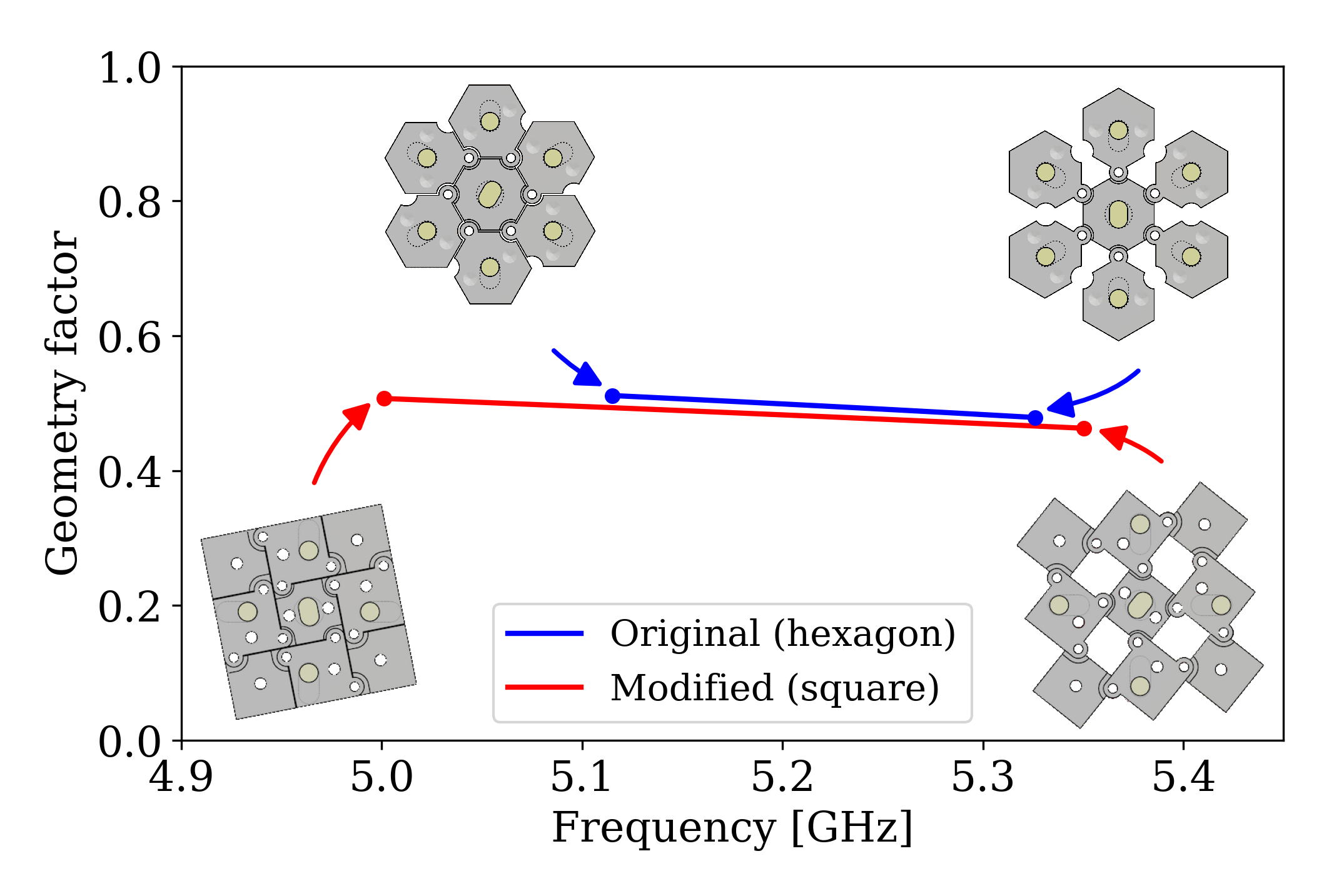}
\caption{Simulated geometry factors over the tunable frequency range for the original (blue) and modified (red) tuning mechanisms.
The endpoints of each curve correspond to the fully compressed and fully extended configurations of the tuning structure.}
\label{fig:comp_hex_quad}
\end{figure}

For the TM$_{020}$ mode of a cylindrical cavity, the electric field profile comprises counter-oscillating components that induce destructive energy storage under a uniform magnetic field (see Eq.~\ref{eq:geo_fac}), leading to a reduced geometry factor ($G=0.15$) compared to the TM$_{010}$ mode ($G=0.69$).
A dielectric structure placed along the cavity center suppresses one of the components, improving the quantity to $\sim$0.5, as illustrated in Fig.~\ref{fig:field_profile}.
Additionally, Ref.~\cite{kim2020exploiting} has demonstrated that the size of the dielectric material affects the resonant frequency.
Our engineered design incorporates an array of thin dielectric rods placed at the center of the cavity. 
These rods are integrated with an auxetic structure that dynamically adjusts their spacing, thereby effectively altering the size of the rod assembly.
The initial tuning mechanism was based on a hexagonal configuration, consisting of six individual blocks hinged to the corners of a central hexagonal block~\cite{bae2024search}.
In this work, we adopt a modified auxetic design featuring a 3$\times$3 square block array arranged in a regular tessellation, adopted from Ref.~\cite{bae2023tunable}.
This design incorporates a thin dielectric rod at the central block and four thicker rods at the side blocks.
Rotation of the central block adjusts the spacing between the rods at equal intervals, effectively increasing (decreasing) the thickness of the dielectric assembly when stretched (compressed).
Compared to the original hexagonal configuration, the square tessellation enables more isotropic deformation while providing larger spacing intervals, thereby improving the frequency tunability, as shown in Fig.\ref{fig:comp_hex_quad}.

\begin{figure}
\centering
\includegraphics[width=\linewidth]{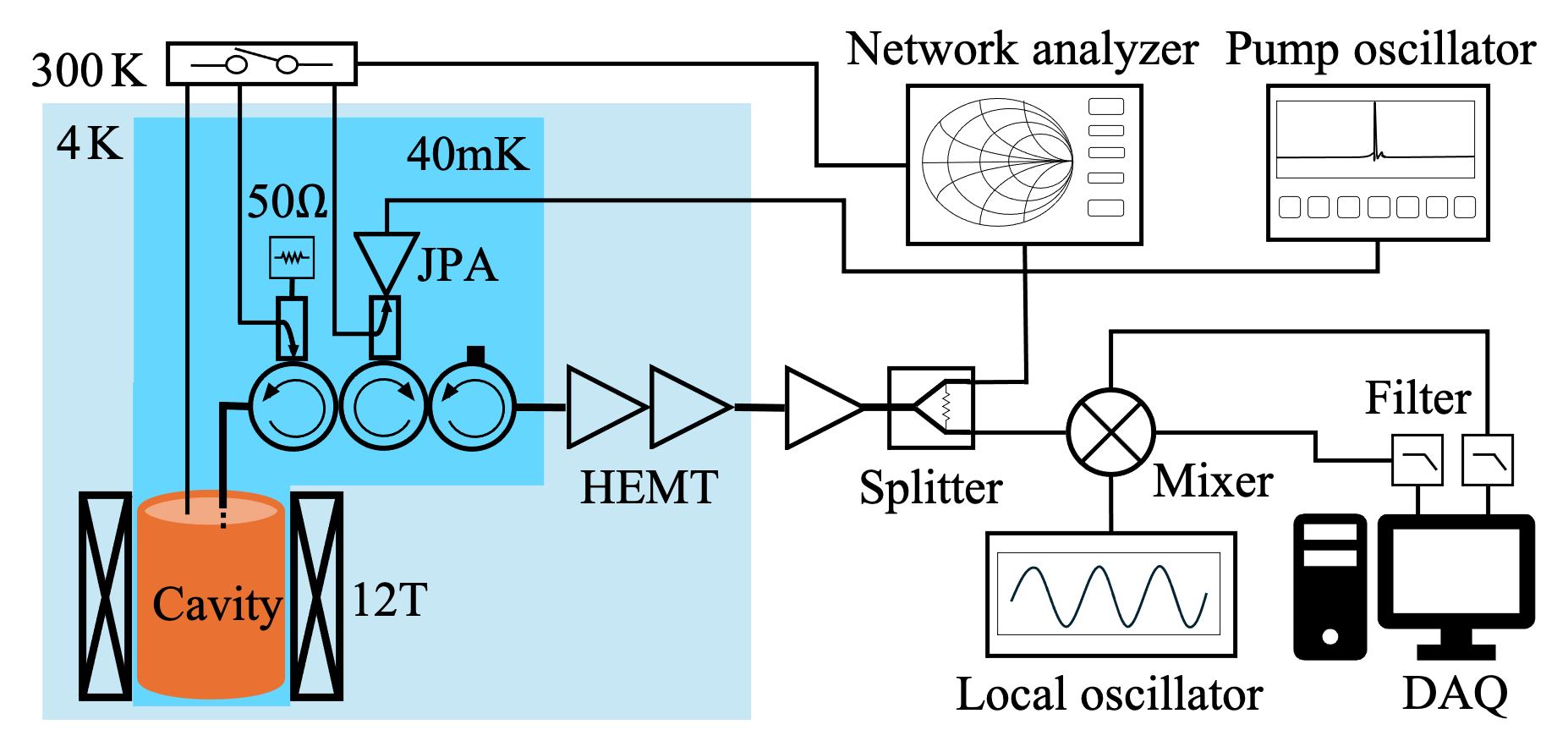}
\caption{Schematic of the experimental setup.}
\label{fig:exp_schem}
\end{figure}

The experiment features a 12-T, Ø96-mm superconducting solenoid and a dilution refrigerator (DR). The solenoid, operating in persistent mode, generates an average 9.8-T field within the cavity, while the DR maintains detector components below 40\,mK. 
The cryogenic system houses the microwave cavity and a readout chain including a Josephson Parametric Amplifier (JPA) and two High Electron Mobility Transistors (HEMTs). 
The signal is further amplified at room temperature, downconverted to an intermediate frequency (IF), digitized, and transformed into the frequency domain for storage. 
A schematic of the haloscope setup is shown in Fig.~\ref{fig:exp_schem}.

The cylindrical cavity, made of 5-mm thick oxygen-free high-conductivity copper, has internal dimensions of $\O78\,{\rm mm} \times 300\,{\rm mm}$, yielding a detection volume of 1.43\,L. 
The tuning rods, fabricated from high-purity dielectric material ($\epsilon=9.7, \tan{\delta} \sim 5\times10^{-6}$), have diameters of 3\,mm and 7\,mm for the central and side rods, respectively.
The side rods feature 2-mm diameter tips on both ends for connection to a pair of tuning structures outside the cavity. 
Each 3$\times$3 auxetic tuning structure, measuring $6.5 \times 6.5 \times 10\,{\rm mm^3}$, is constructed from aluminum, with each block thermally linked to the cavity via thin copper wire for effective cooling of the tuning rods.
A piezoelectric actuator, coupled to the central rod via a 100:24 gear system, drives rotational tuning of the auxetic structure.
Optimized through finite element simulations~\cite{comsol}, the rods have an initial 6.5-mm spacing and a 2.7-mm travel length, enabling frequency tuning from 5.00 to 5.29\,GHz.
The simulation study also estimate a form factor of approximately 0.5 throughout this range.
The unloaded cavity quality factors were measured to be $\sim$110,000, remaining relatively uniform over the scan.

Our flux-driven JPA, similar to the design in Ref.~\cite{kutlu2021characterization}, consists of a $\lambda/4$ coplanar waveguide terminated with a DC Superconducting Quantum Interference Device.
To protect it from external magnetic fields, the JPA is enclosed within a three-layer shield made of aluminum, cryoperm, and NbTi alloys.
Its characteristics are dictated by a magnetic flux that modulates the nonlinear inductance of the Josephson junctions, while a microwave pump signal enables parametric amplification via three-wave mixing.
Tuned by a superconducting coil, the JPA operates over the 5.00--5.32\,GHz range.
The JPA was operated in a phase-insensitive mode by detuning its resonance from the cavity frequency by 200\,kHz.

The optimal operating conditions of the JPA were determined using the Nelder-Mead (NM) algorithm~\cite{nelder1965simplex}, a heuristic method that iteratively adjusts two parameters—flux and pump power—to find the minimum noise temperature~\cite{kim2024parameter}.
The algorithm initializes these parameters based on pre-scanned resonance and gain profiles.
At each iteration, a new test point is selected within the two-dimensional parameter space, and the noise temperature is estimated using the Noise Visibility Ratio (NVR)~\cite{Friis:ProcIRE:1944, PhysRevX.5.041020}, defined as:
\begin{equation}
    {\rm NVR} = \frac{P^{\rm on}}{P^{\rm off}} = \frac{G_{{\rm JPA}} T ^{{\rm on}}_{{\rm sys}} }{T ^{{\rm off}}_{{\rm sys}}},
\end{equation}
where the superscripts denote the pump on and off, respectively, and $G_{\rm JPA}$ is the JPA gain determined using a network analyzer.
NVR-based estimation, validated against the Y-factor method, enables time-efficient in-situ noise evaluation when integrated with the NM algorithm.
The number of iterations was set to 10, which was sufficient to find the local minimum noise temperature, typically requiring 1 minute.
The system noise was estimated to be as low as 550\,mK over its tunable range with a typical gain of 20\,dB.

The data acquisition (DAQ) sequence was as follows.
The cavity was characterized using a network analyzer (NA): the resonant frequency and $Q_l\approx35,000$ from transmission measurements and $\beta\approx2.0$ from Smith circle fitting~\cite{IEEE:Q_coupling:1984}.
To operate in phase-insensitive mode, the JPA resonance was detuned by approximately 200\,kHz from the cavity resonance.
The noise temperature was estimated using the NM algorithm by optimizing flux bias and pump power.
Signals were extracted via a strongly coupled antenna and amplified sequentially by the JPA, a pair of cryogenic HEMTs, and a room-temperature HEMT.
Every 10 seconds, the acquired signal was down-converted to a 3-MHz intermediate frequency—a quiet region in the ambient noise spectrum—using a local oscillator and an IQ mixer.
The two quadrature components were processed separately up to digitization at 20\,MHz, then merged in software.
The time-series data was Fourier transformed over a 1-MHz span with a resolution bandwidth of 100\,Hz before stored for offline analysis.
DAQ paused every 6 minutes for a JPA sanity check, rerunning the NM algorithm if the gain deviated by more than 1\,dB from the prior measurement.
Data was collected for 30--120 minutes per tuning step depending on the JPA performance, with a DAQ efficiency of 87.2\%.
The cavity was then tuned by 50\,kHz, approximately one-third of its loaded bandwidth, to the next search frequency.
The experiment ran from August 22th, 2024 to January 21th, 2025 with several interruptions due to facility maintenance, yielding a total data collection period of 152 days.

The data analysis followed the procedure outlined in~\cite{bae2024search}. 
For each 10-second raw spectrum, the baseline shape--affected by factors like impedance mismatch--was removed using a Savitzky-Golay (SG) filter~\cite{savitzky1964smoothing}, and the spectrum was normalized to this baseline.
The pre-processed spectrum was rescaled to ensure consistent scaling across different frequency ranges, using the ratio $(P_{{\rm sig}} / \delta P_{{\rm sys}})^{-1}$, where $P_{{\rm sig}}$ is the expected signal power and $\delta P_{{\rm sys}}$ is the mean noise power, to reflect the frequency dependence of the cavity response.
In the rescaling, the expected signal power was estimated assuming KSVZ axions with $\rho_a = 0.45~\rm GeV/cm^3$~\cite{10.1093/mnras/stae034,PhysRevD.99.023012,deSalas_2021}.
The rescaled spectra were vertically combined using inverse-variance weighting to yield a unified single spectrum spanning 5.07--5.17\,GHz.
For each frequency bin, power excesses in subsequent bins were weighted according to the Standard Halo Model for axions~\cite{turner1990periodic} and combined.
Repeating this over the full span produced a grand spectrum, with each bin representing its maximum-likelihood estimate.

The SG filter, applied for baseline removal, smooths the power excess and introduces correlations that influence the signal-to-noise ratio (SNR) in the final results~\cite{JHEP.115.11.2023}.
The performance of the SG filter was evaluated through a comprehensive Monte Carlo study.
The filter parameters, polynomial degree and window size, were optimized to maximize the SNR in the presence of axion signals. 
Synthetic axion signals with an expected ${\rm SNR} = 5.0$ were overlaid onto raw baseline spectra, and the full analysis procedure was repeated 400 times to determine the optimal filter configuration, found to be degree 4 and window size 901.
This process was carried out at 10 different frequencies, yielding an overall signal reconstruction efficiency of $\epsilon_{\rm sig}=79.5\pm2.5\%.$
The corresponding noise spectrum followed a normal distribution $\mathcal{N}(0.00,0.88)$, resulting in an effective SNR of $4.59\pm0.13$.

A statistical hypothesis test was employed to evaluate the presence of an axion-induced signal in the grand spectrum.
The spectrum was re-normalized to unit variance to facilitate sensitivity analysis.
For axions with ${\rm SNR} = 5.0$, a detection threshold of 3.718 was set, corresponding to a 90\% confidence level, to identify potential signal candidates.
Within the scanned range, 12 candidates exceeded the threshold and were subjected to further investigation.
Subsequent rescans led to all excesses falling below the threshold, indicating them as statistical fluctuations.

\begin{figure*}
\centering
\includegraphics[width=0.95\linewidth]{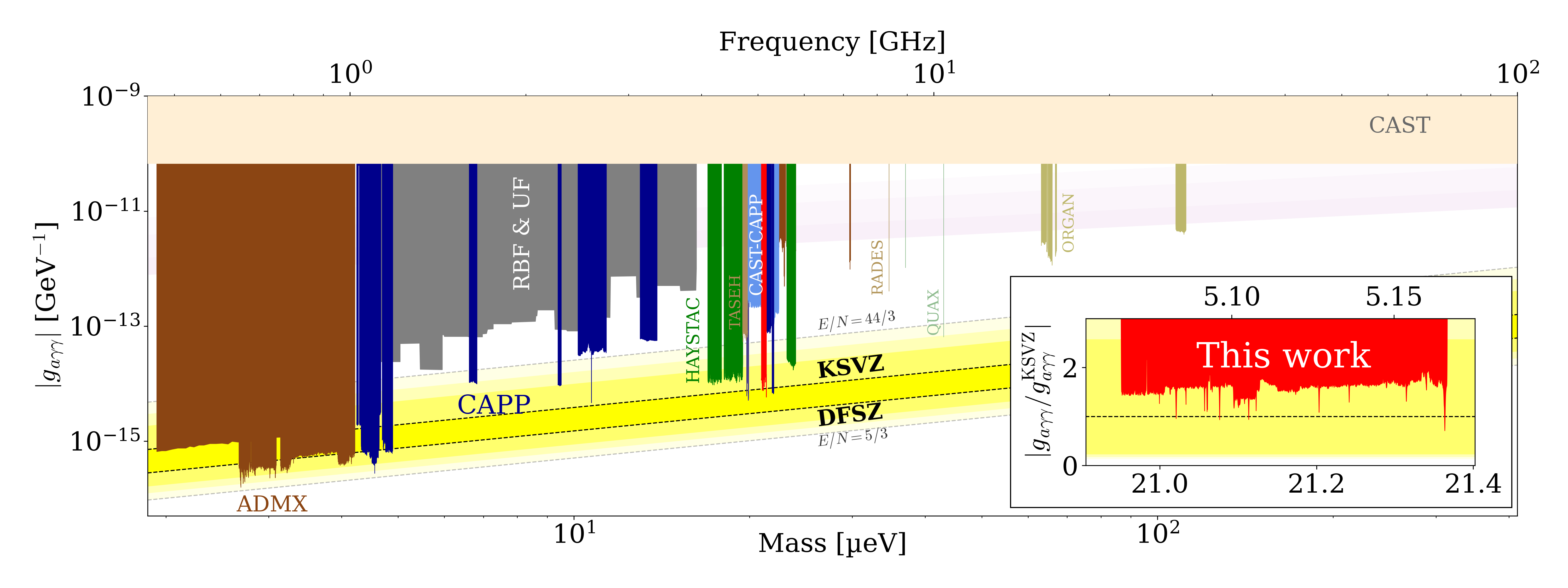}
\caption{Exclusion limits on axion–photon coupling as a function of axion mass from cavity haloscope searches: RBF~\cite{PhysRevLett.59.839,PhysRevD.40.3153},UF~\cite{PhysRevD.42.1297,HAGMANN1996209}, ADMX~\cite{PhysRevLett.104.041301,PhysRevLett.120.151301,PhysRevLett.124.101303,PhysRevLett.127.261803,PhysRevLett.121.261302,PhysRevLett.134.111002,10.1063/5.0122907}, CAPP~\cite{PhysRevLett.124.101802,PhysRevLett.125.221302,PhysRevLett.126.191802,PhysRevLett.128.241805,PhysRevD.106.092007,PhysRevLett.130.091602,PhysRevLett.130.071002,PhysRevLett.131.081801,PhysRevLett.133.051802,PhysRevX.14.031023}, HAYSTAC~\cite{PhysRevLett.118.061302,Backes:2021aa,PhysRevD.97.092001,PhysRevD.107.072007,PhysRevLett.134.151006}, TASEH~\cite{PhysRevLett.129.111802}, CAST-CAPP~\cite{Adair:2022aa}, RADES~\cite{Alvarez-Melcon:2021aa}, QUAX~\cite{PhysRevD.99.101101,PhysRevD.103.102004,PhysRevD.106.052007,PhysRevD.108.062005,PhysRevD.110.022008}, ORGAN~\cite{MCALLISTER201767,doi:10.1126/sciadv.abq3765,PhysRevLett.132.031601}.
The dashed black lines indicate the KSVZ and DFSZ model predictions with the yellow band representing theoretical uncertainties.
The newly excluded region by this work is shown in red and magnified in the inset where the coupling is normalized to the KSVZ model.}
\label{fig:exclusion}
\end{figure*}

Since the power excess was normalized in units of the KSVZ axion signal during rescaling, the standard deviation $\sigma$ in each bin quantifies statistical fluctuations and corresponds to the inverse of the SNR for the KSVZ axion coupling strength.
Accordingly, the null hypothesis was defined as an axion-induced excess corresponding to a coupling of $\sqrt{5\sigma/\epsilon_{\rm sig}}\times g^{\rm KSVZ}_{a\gamma\gamma}$.
The average $\sigma$ across a total of 1,014,602 scanned frequency bins was found to be 0.449.
Since no excess was observed, the null hypothesis was rejected at the 90\% CL, setting an upper limit on the axion-photon coupling of $g_{a\gamma\gamma} \gtrsim 1.7 \times g^{\rm KSVZ}_{a\gamma\gamma}$ over the mass (frequency) range 20.95--21.37\,$\mu\rm{eV}$ (5.07--5.17\,GHz), as shown in Fig.~\ref{fig:exclusion}.

The final results accounted for both statistical and systematic uncertainties. 
The system noise temperature exhibited statistical fluctuations of approximately 30\,mK throughout the experiment. 
The statistical and fitting uncertainties in measuring the loaded quality factor, were 1.1\% and 0.6\%, respectively. 
The coupling coefficient, determined from Smith circle fitting, had a typical error of 4.2\%. 
The uncertainty in the geometry factor was evaluated through numerical simulations considering potential misalignment of the tuning rods due to the complex tuning structure. 
Each rod was randomly displaced by up to 50\,$\mu$m, and the geometry factor was computed for each configuration. 
From 1,000 simulations, statistical fluctuations in the factor remained below 0.1\% across the scanned frequency range.

In conclusion, we have conducted a high-sensitivity axion search near 21\,$\mu$eV using the TM$_{020}$ mode of a cylindrical cavity equipped with an innovative tuning mechanism.
Our results achieved near-KSVZ sensitivity over a 100-MHz range, marking a significant advancement in axion searches at higher masses.
By successfully employing a higher-order resonant mode, traditionally considered challenging for both sensitivity and tunability, this work demonstrates its viability for probing axions in previously unexplored mass regions.
These findings not only set new constraints on the axion parameter space but also lay the groundwork for future experimental efforts targeting higher-mass axion dark matter.

\begin{acknowledgments}
This work was supported by the Institute for Basic Science (IBS-R017-D1-2024-a00 and IBS-R040-C1-2025-a00) and by JSPS KAKENHI (Grant No.~JP22H04937).
J. Jeong was partially supported by the Knut and Alice Wallenberg Foundation.
Y. Kim was partially supported by the Alexander von Humboldt Foundation.
A. F. Loo was supported by a JSPS Postdoctoral Fellowship.
J. E. Kim was partially supported by the Korea National Science Foundation.
\end{acknowledgments}

\bibliographystyle{apsrev4-2}
\bibliography{main}

\end{document}